\def\ket#1{{}\mid #1\rangle{}}

\def\obraket#1#2#3{{}\langle#1\mid #2\mid #3\rangle{}} 

\documentclass[12pt]{article}

\begin{document}

\vskip 2cm
\title{\bf On the consequences of the fact that atomic levels have a certain width}

\author{Claude Billionnet\\
Centre de Physique Th\'eorique, \'Ecole Polytechnique, CNRS,\\ 91128 Palaiseau, France\\
claude.billionnet@cpht.polytechnique.fr}

\maketitle

\begin{abstract}

This note presents two ideas. The first one is that quantum theory has a fundamentally perturbative basis but leads to nonperturbative states which it would seem natural to take into account in the foundation of a theory of quantum phenomena. The second one consists in questioning the validity of the present notion of time. Both matters are related to the fact that atomic levels have a certain width. This note is presented qualitatively so as to evidence its main points, independently of the models on which these have been tested.
\end{abstract}


\section{ Introduction}

The fact that optical spectral lines have a certain width is fundamental for Quantum Mechanics. Without such a width, Quantum Mechanics could not have been discovered since atoms would not have emitted any light. (And besides no life on Earth would be possible.) It is therefore remarkable that the theory does not take them into account {\sl ab initio}. Clearly, one may say that it is an approximate theory. But usually if a theory is often developed through a simplification of reality, neglecting irrelevant phenomena, the main phenomenon is, of course, described by the theory. Here, the root of the theory deals with the emission of light by atoms (with many consequences independent of this phenomenon) but this emission can only be described through the introduction of a perturbation.
Quantum Mechanics is thus a kind of zero-order description and the approach is a perturbative one. Its well-known successes are due to the small effects this perturbation has. However one may feel uncomfortable with respect to the method. A way of pointing out that we have here the zero-order of something more complex is to say that the Ritz-Rydberg law (the fact that the frequencies of the emitted light are differences) can be stated for numbers, the abscissas of the line-shapes' peaks, but not for the more complex objects the line-shapes are, the latter also depending on the environment.

So as to overcome this defect one might dream of a theory which would describe the emission of radiation by other means than through the perturbation of a theory (ordinary Quantum Mechanics) in which this emission is impossible. In this note, we shall give a more precise reason for trying to build a theory which would not be perturbative in the foregoing sense (Sec. 2). We also mention (Sec. 3) that the line widths are likely to entail a reconsidering of the familiar notion of time. 

\section{ Necessary mixing of matter and radiation and nonperturbative states}

\subsection{Existence of two types of resonances}

In the description of atomic states in the zero order, matter and the transverse radiative field are well separated. But in fact, very often, this separation is impossible. This is the case, for instance, when an atom (assumed to have two levels only, for the sake of simplicity) is situated inside a cavity whose walls are perfect mirrors, the cavity being tuned to the transition between the two atomic levels. In the simplest model describing the coupling of the atom with one mode of the radiation in the cavity (Jaynes-Cummings model [1]), the atom-field system's Hamiltonian has two eigenstates  $\ket +$ and $\ket -$, in which the two atom-states $\ket f$ (fundamental) and $\ket e$ (excited) are mixed with the states $\ket \gamma$ (one photon) and $\ket\Omega$ (vacuum) of the field in the cavity. (We restrict ourselves to considering only one photon.) We have
$$
\ket\pm=2^{-1/2}\big(\ket e\otimes \ket\Omega\pm\ket f\otimes\ket\gamma
\big)
$$
(There are other eigenstates, including $\ket f\otimes\ket \Omega$.)
When the mirrors are no longer perfect, energies $E_\pm$ of the foregoing states are changed into two complex numbers, whose imaginary parts are interpreted as the two levels' widths and related to the states' life-times: these numbers are poles of matrix elements of the resolvent of the atom-field system's Hamiltonian; these poles are called resonances. It can be shown that one of them is the trace (as defined hereafter) of the upper state of the atom, whereas the other one is not; the latter may be related to the energy of the photon inside the cavity [2].

This example illustrates a very general phenomenon: one can expect that every coupling of some matter (system ${\cal S}$) to radiation yields such a set of resonances, which in a way represent mixed states.  (We give a summary justification of this statement in Sec. 2.3.) Resonances which are not the `trace' of atomic states are those we consider as nonperturbative; we call them nonstandard henceforth. In the above example, one of the resonances is therefore standard and the other one nonstandard. Let us make precise what we mean by `trace'. In the coupling, several physical parameters are involved. In the example, three can be identified: the strength of the atom-cavity coupling (governed by the dipole moment of the atomic transition and the fine structure constant $\alpha$), the mirrors' reflectivity, and possibly an atom-cavity detuning. When these parameters vary, the resonances move. (They may become real. They are then eigenvalues of the {\it global} Hamiltonian. In this note, the term `resonance' will generically encompass this case.) When the matter-radiation coupling tends to $0$, the other parameters keeping their values, a resonance may tend to an eigenvalue of $H_{\cal S}$, the Hamiltonian of ${\cal S}$ alone. We call it `standard', and those who do not have that property are called nonstandard. It is in this sense that standard resonances are considered as the trace of the eigenvalues of ordinary Quantum Mechanics. This notion of a trace is therefore of a perturbative origin and we are now going to give reasons for diminishing its pertinence.

\subsection{ Reasons for not dissociating the two types of resonances}

Besides the existence of nonstandard resonances (numerous, in fact, as will be seen later on) there is another point we would like to highlight~: the `standard'/`nonstandard' distinction stems from the choice of a quite specific path in the parameter space. If we do not consider this zero-coupling limit, that is to say eventually, if we do not take a perturbative point of view, we are faced with all the resonances and there is {\sl a priori} no reason to give greater importance to some of them. This is the main point of the paper. This position is clearly justified in the cavity case: it cannot be denied that both states are similar and the same is true for the resonances into which the energies of these `states' transform when the reflexivity changes from $1$ to a neighbouring value.

The importance of the path for the standard vs. nonstandard labeling of the resonances shows up in the following fact. Along a (real) closed path in the parameter space, two resonances may interchange; in particular, a standard resonance may change into a nonstandard one and conversely. A possibility related to this fact is that two resonances may coincide; in particular, a standard resonance may coincide with a non-standard one. Still another way of underlining the importance of the paths is to notice that a standard resonance may change into a nonstandard one through a variation of the value of one of the parameters (not the coupling strength, of course). This issue is studied in [3].

For the reasons mentioned so far, it seems to us that one resonance type must not be {\sl a priori} stressed upon to the detriment of the other. The distinction standard/nonstandard resonance does not seem to have any physical meaning, fundamentally, even if it has the above-mentioned mathematical meaning. All the resonances should be considered as similar.  An approach of the resonances as poles of the $S$ matrix, when it can be defined, would lead to the same conclusion.

Incidentally, and although we do not want to insist upon this point here, it has to be noted that the total number of resonances is far greater than the number of eigenstates of $H_{\cal S}$. In the case of the atom inside a perfect cavity, this is related to the fact that such eigenstates as $\ket e\otimes\ket{n \gamma}\pm \ket f\otimes\ket{(n+1)\gamma}$ are also to be considered and that they have energies which differ from $E_{\pm}$. In the general case, it is the infinite dimension of the space of radiation states (number of bosons and dimension of the one-boson state space) which is responsible for the infinity of the resonances. This is to be related to the fact that a hermitian operator in $E\otimes F$ has $m\times n$ eigenvalues if $\rm{dim}(E)=m$ and $\rm{dim}(F)=n$, although the multivaluedness of the functions involved in our case makes things more complicated. When concerned with the coupling of a discrete-level system with a continuum, things are often presented differently: focus is on what the eigenvalues of the Hamiltonian at zero coupling become when a small coupling is turned on (see for instance [4], for a rigorous analysis), instead of directly considering all the resonances at non-zero coupling. The former point of view is the perturbative standpoint we want to underline and challenge.

\subsection{Justification of the existence of nonstandard resonances}

Let us show how the existence of nonstandard resonances, illustrated in the atom-cavity case, can be seen to be something very general in the coupling of matter with radiation. The generalization from the atom-cavity case can be performed in two steps. First we consider the atom-photon interaction in the whole space, then every coupling of a discrete-level system with a continuum, an instance being the (quark-antiquark)-gluon interaction. The existence of nonstandard resonances in the case where the atom is in the whole space can be inferred in the following way. Decreasing the mirrors' reflectivity, one expects the resonances to move without disappearing. When the Wigner-Weisskopf's Hamiltonian for the matter-radiation coupling is used (see (6) of [2] for its expression), and if $k$ denotes the wave-number of a radiative boson, non-standard resonances are caused by poles of the coupling function $g(k):=\obraket {f,k}H{e,\Omega}$. It is a good approximation to treat the atom-cavity coupling by taking a Lorentzian function for $g$, with a small width when the mirrors are good. Therefore the better the quality of the mirrors the smaller the imaginary part of the pole of $g$. When the mirrors are removed, the coupling function still has poles, even if they may have moved away to infinity, as it can be seen to be the case when ${\cal S}$ is a particle in a square well (ref. 3 of [5]). In a model taking into account only two levels of the Hydrogen atom coupled to the transverse electromagnetic field, poles of $g$ are at a finite distance [6] (a distance related to the extension in space of the eigenstates corresponding to the two levels). These poles are the origin of the nonstandard resonances [2] since for small couplings these resonances are close to the poles of $g$. For the foregoing atom, nonstandard resonances have no physical effect [2]. For a strong interaction, a rough model yields analogous poles and resonances the effect of which may be noticeable [5]. (As regards this strong interaction, another continuum has been  considered by some authors [7,8]: that of meson-meson states, and nonstandard resonances have indeed been studied in that context; they are called `continuum poles', whereas those related to the $q\overline q$ system alone are called `confinement poles'.)

Nonstandard resonances have been studied elsewhere by the author on various examples (see for instance [5] and references therein). They are also present in works by other authors in various contexts. Apart from the above-mentioned meson-meson case , [9] (Chapter 13) and [10] deal with the case where the radiation field is replaced by the $\pi$ meson's field. There is also a great amount of mathematical literature devoted to the study of the fundamental state of a system ${\cal S}$ coupled to a massless field or to the determination of the S matrix, in various models. Let us quote for instance [11, 12, 13, 14]. This is related to our issue since resonances are poles of the S matrix and such a fundamental state may come from what we call a nonstandard resonance, for instance when the coupling is large enough.

To sum up, the coupling of ${\cal S}$ with the continuum of the radiation states is therefore responsible for both the widening of the levels of ${\cal S}$ (standard resonances) and the existence of nonstandard resonances, both accounting for a more or less significant mixing of the matter with the radiation. We would like the description of the interaction to give a well-balanced weight to both types of resonances. Let us now see how the matter-radiation coupling seems to question the very notion of time.

\subsection{Lack of physical basis for the notion of time}

The time standard is presently defined through the frequency of an atomic transition between two practically stable levels. For atomic levels considered as infinitely narrow, a transition between two levels, which can only be imagined, indeed yields a photon of (precise) frequency $\nu$ and, by means of an appropriate device, a wave with the same frequency. The latter enables us to mark a sequence of instants defining a time standard. In order to get a smaller time unit, on may think of using the strong interaction, but the coupling with the radiation (now the gluon) puts us in the situation described in Section 2. The relevant resonances are then numerous. If we assume that it is still possible to associate physical states to them, these are mixed states. The widths are large or small, depending on the resonances taken into account. (Note that nonstandard resonances may have small widths [5].) Moreover, the resonances depend on the radiative environment. Therefore it becomes difficult to imagine producing an object with a well-defined frequency. This seems to question the physical basis of the current notion of time.

\section{Conclusion}

\smallskip
We saw how the traditional-level width (imaginary part of the resonances we call standard) leads to a nonperturbative viewpoint consisting in {\sl a priori} taking into account the particular `states' of the matter-radiation system associated with all the resonances of the system's Hamiltonian, rather than the only levels of the decoupled $\cal S$. This would likely lead to no longer separate the matter and radiation fields in the global system's Hamiltonian, on the assumption that a description by a Hamiltonian remains possible. We indeed saw that the width of the traditional levels has also an impact on the very notion of time.

We are well aware that we here suggest to give up a perturbative standpoint without giving clues for doing it effectively. The matter-radiation mixing leaves us a bit powerless. As regards being deprived of the usual notion of time, it is perhaps even worse. Nevertheless, the physical situation does seem to be as described.

\end{document}